%% file: main.tex
\documentclass[sigconf]{acmart}
\AtBeginDocument{%
  }

\newcommand{\cameraready}{}

\copyrightyear{2026}
\acmYear{2026}
\setcopyright{cc}
\setcctype{by}
\acmConference[ICSE '26]{2026 IEEE/ACM 48th International Conference on Software Engineering}{April 12--18, 2026}{Rio de Janeiro, Brazil}
\acmBooktitle{2026 IEEE/ACM 48th International Conference on Software Engineering (ICSE '26), April 12--18, 2026, Rio de Janeiro, Brazil}
\acmPrice{}
\acmDOI{10.1145/3744916.3773160}
\acmISBN{979-8-4007-2025-3/26/04}

\input{chaps/main-defs.tex}

\begin{document}

\input{chaps/00-titleauthor}
\input{chaps/00-abs}

\input{chaps/00-info}

\ifdefined\cameraready
\else
\setcopyright{none}
\settopmatter{printacmref=false}
\renewcommand\footnotetextcopyrightpermission[1]{\footnote{Preprint. Accepted by ICSE 2026.}}
\fi

\maketitle

\input{chaps/01-intro}

\input{chaps/02-motivation-example}

\input{chaps/03-method}

\input{chaps/04-experiments}

\input{chaps/05-discussion}
\input{chaps/06-related-work}
\input{chaps/07-conclusion}


\begin{acks}
This research is supported in part by the National Natural Science Fundation of China (62572300), the Minister of Education, Singapore (MOE-T2EP20124-0017, MOET32020-0004), the National Research Foundation, Singapore and the Cyber Security Agency under its National Cybersecurity R\&D Programme (NCRP25-P04-TAICeN), DSO National Laboratories under the AI Singapore Programme (AISG Award No: AISG2-GC-2023-008-1B), and Cyber Security Agency of Singapore under its National Cybersecurity R\&D Programme and CyberSG R\&D Cyber Research Programme Office. Any opinions, findings and conclusions or recommendations expressed in this material are those of the author(s) and do not reflect the views of National Research Foundation, Singapore, Cyber Security Agency of Singapore as well as CyberSG R\&D Programme Office, Singapore.
\end{acks}

\bibliographystyle{ACM-Reference-Format}
\bibliography{main}


\end{document}

%% file: chaps/main-defs.tex
\usepackage{listings}
\usepackage{xspace}
\usepackage{xcolor}
\usepackage{subfig}
\usepackage{enumitem}
\usepackage{enumitem}
\usepackage{multirow}
\usepackage{pifont}
\usepackage{framed}

\definecolor{tableau1}{RGB}{31,119,180}
\definecolor{tableau2}{RGB}{255,127,14}
\definecolor{tableau3}{RGB}{44,160,44}
\definecolor{tableau4}{RGB}{214,39,40}
\definecolor{tableau5}{RGB}{148,103,189}
\definecolor{tableau6}{RGB}{140,86,75}
\definecolor{tableau7}{RGB}{227,119,194}
\definecolor{tableau8}{RGB}{127,127,127}
\definecolor{tableau9}{RGB}{188,189,34}
\definecolor{tableau10}{RGB}{23,190,207}

\definecolor{codegreen}{rgb}{0,0.6,0}
\definecolor{codegray}{rgb}{0.5,0.5,0.5}
\definecolor{codepurple}{rgb}{0.58,0,0.82}

\definecolor{shadecolor}{RGB}{240,240,240}

\lstdefinestyle{javastyle}{
	commentstyle=\color{codegreen},
	keywordstyle=\color{magenta},
	numberstyle=\tiny\color{codegray},
	stringstyle=\color{codepurple},
	basicstyle=\ttfamily\footnotesize,
	breakatwhitespace=false,
	breaklines=true,
	captionpos=b,
	keepspaces=true,
	numbers=left,
	numbersep=5pt,
	showspaces=false,
	showstringspaces=false,
	showtabs=false,
	tabsize=2
}

\newcommand{\techname}{MINES\xspace}

\newcommand{\finding}[1]{
\begin{snugshade}
\noindent{#1}
\end{snugshade}
}

\newcommand{\trainticket}{\textit{Train-Ticket}\xspace} 
\newcommand{\nicefish}{\textit{NiceFish}\xspace}
\newcommand{\nextcloud}{\textit{NextCloud}\xspace}
\newcommand{\gitea}{\textit{Gitea}\xspace}
\newcommand{\mastodon}{\textit{Mastodon}\xspace}

\ifdefined\cameraready
\newcommand{\mytablefontsize}{\small}

\newcommand{\motivatingwidth}{.85\textwidth}

\else
\newcommand{\mytablefontsize}{}

\newcommand{\motivatingwidth}{.95\textwidth}

\fi

\newcommand{\logsn}{\textit{ManualNorm}\xspace}
\newcommand{\logln}{\textit{LLMNorm}\xspace}
\newcommand{\logsa}{\textit{ManualAbnormal}\xspace}
\newcommand{\logia}{\textit{InjectAbnormal}\xspace}

%% file: chaps/00-titleauthor.tex
\title{\techname: Explainable Anomaly Detection through Web API Invariant Inference}

\author{Wenjie Zhang}
\authornote{This work was partially conducted when Wenjie Zhang was visiting Shanghai Jiao Tong University}
\email{wjzhang@nus.edu.sg}
\orcid{0000-0002-2669-1837}
\affiliation{%
  \institution{National University of Singapore}
  \country{Singapore}
}

\author{Yun Lin}
\authornote{Yun Lin is the corresponding author.}
\email{lin_yun@sjtu.edu.cn}
\affiliation{%
  \institution{Shanghai Jiao Tong University}
  \city{Shanghai}
  \country{China}
}

\author{Kwok Chun Fung Amos}
\email{e1373883@u.nus.edu}
\affiliation{%
  \institution{National University of Singapore}
  \country{Singapore}
}

\author{Xiwen Teoh}
\email{xiwen.teoh@u.nus.edu}
\affiliation{%
  \institution{National University of Singapore}
  \country{Singapore}
}

\author{Xiaofei Xie}
\email{xfxie@smu.edu.sg}
\affiliation{%
  \institution{Singapore Management University}
  \country{Singapore}
}

\author{Frank Liauw}
\email{Frank_LIAUW@tech.gov.sg}
\affiliation{%
  \institution{GovTech}
  \country{Singapore}
}

\author{Hongyu Zhang}
\email{hongyujohn@gmail.com}
\affiliation{%
  \institution{Chongqing University}
  \city{Chongqing}
  \country{China}
}

\author{Jin Song Dong}
\email{dcsdjs@nus.edu.sg}
\affiliation{%
  \institution{National University of Singapore}
  \country{Singapore}
}

\renewcommand{\shortauthors}{Zhang et al.} 

%% file: chaps/00-abs.tex
\begin{abstract}
    Detecting the anomalies of web applications,
    important infrastructures for running modern companies and governments,
    is crucial for providing reliable web services.
    Many modern web applications operate on web APIs (e.g., RESTful, SOAP, and WebSockets),
    their exposure invites intended attacks or unintended illegal visits,
    causing abnormal system behaviors.
    However, such anomalies can share very similar logs (sometimes even identical logs) with normal logs,
    missing crucial information (which could be in database) for log discrimination.
    Further, log instances can be also noisy,
    which can further mislead the state-of-the-art log learning solutions to learn spurious correlation,
    resulting superficial models and rules for anomaly detection.

    In this work, we propose \techname
    which infers explainable API invariants for anomaly detection from the \textit{schema level}
instead of detailed raw log instances,
    which can
    (1) significantly discriminate noise in logs to identify precise normalities and
    (2) detect abnormal behaviors beyond the instrumented logs (e.g., regarding the database state or session state).
    Our learned invariants can capture API preconditions such as
    (1) ``\textit{what is the legitimate database state to initiate the call events?}'' and
    (2) ``\textit{what are the constraints to satisfy between different API calls?}''.
    Then
    we translate the invariants into executable Python code to verify its consistency with the runtime logs.
    Technically, \techname
    (1) converts API signatures into table schema to enhance the original database shema; and
    (2) infers the potential database constraints (such as reference constraint and check constraints) on the enhanced database schema to capture the potential relationships between APIs and database tables.
    \techname uses LLM for extracting potential relationship based on two given table structures;
    and use normal log instances to reject and accept LLM-generated invariants.
    Finally, \techname translates the inferred constraints into invariants to generate Python code for verifying the runtime logs.
    We extensively evaluate \techname on web-tamper attacks on the
    benchmarks of \trainticket, \nicefish, \gitea, \mastodon, and \nextcloud against baselines such as LogRobust, LogFormer, and WebNorm.
    The results show that \techname achieves high recall (more than 14\% over LogRobust, LogFormer, and WebNorm) for the anomalies while introducing almost zero false positives,
    indicating a new state-of-the-art.
\end{abstract}

%% file: chaps/00-info.tex
\ccsdesc[500]{Security and privacy~Web application security}
\ccsdesc[500]{Computing methodologies~Anomaly detection}
\ccsdesc[300]{Software and its engineering~Software testing and debugging}
\ccsdesc[300]{Information systems~Web services}
\ccsdesc[100]{Computing methodologies~Natural language processing}

\keywords{Log-based anomaly detection, Specification mining, API analysis, Web application, Database}



%% file: chaps/01-intro.tex
\section{Introduction}

Web applications are crucial infrastructures for running companies and governments in modern society \cite{hoffman2024web,neumann2018analysis,qi2022correlation}.
Many applications operate through web APIs (e.g., RESTful and WebSockets) exposed to the public,
which can attract intentional attacks or unintended illegal access,
resulting in approximately 35\% of abnormal system behaviors, according to a Salt Security report \cite{salt2023api}.
Therefore, detecting such anomalies is important for operating and maintaining reliable web services \cite{he2021survey}.
To this end, researchers and industry practitioners have developed a variety of automated log-based anomaly detectors \cite{ye2024spurious}.
Traditional log anomaly detection approaches rely on predefined rules \cite{hansen1993automated,oprea2015detection,prewett2003analyzing,rouillard2004real,roy2015perfaugur,yamanishi2005dynamic,yen2013beehive}, which are limited to specific application scenarios and require domain expertise \cite{du2017deeplog}.
In recent years, researchers have proposed learning-based approaches to automatically learn normal behaviors from logs,
which can be categorized into two types:
\begin{itemize}[leftmargin=*,topsep=2pt]
  \item \textbf{Model-learning Approaches.}
    Researchers have proposed detecting abnormal logs by training deep learning models in a supervised or unsupervised manner \cite{du2017deeplog,huang2020hitanomaly,le2021log,zeng2021watson,meng2019loganomaly,yang2021plelog}.
    In supervised learning solutions,
    anomaly detection is reduced to a binary classification problem \cite{du2017deeplog,huang2020hitanomaly,le2021log,zeng2021watson}.
    In contrast, unsupervised learning solutions \cite{meng2019loganomaly,yang2021plelog} learn
    normalities from collected normal logs,
    reporting logs as anomalies if they deviate from these normalities based on predefined metrics.
    However, both solutions can suffer from limited explainability \cite{haar2023analysis} and
    the distribution shift problem \cite{liu2021towards}.
  \item \textbf{Rule-learning Approaches.}
    Facilitated by advances in LLMs, recent works \cite{liao2024detecting} learn project-specific rules in the form of first-order logic from collected logs.
    The learned rules are used to check against runtime logs.
    While such a solution can improve both the explainability of anomalies and detection accuracy for web-tampering attacks \cite{liao2024detecting},
    it still suffers from generating false positive rules due to abundant log noise and producing false negatives when crucial information is unavailable in the logs (see Section~\ref{sec:motivating-example}).
\end{itemize}

State-of-the-art approaches have fundamental limitations
in learning normalities solely from raw log instances,
which presents challenges in log observability and log noise discrimination.
Specifically, the more comprehensive logs we collect,  
the more informative we can potentially learn a discriminative model.
However, the logs can also be more noisy to mislead the model to capture \textit{spurious correlation} rather than causality.

For some APIs, comprehensive logging is necessary to learn true normalities.
For example, a log event indicating the \textit{successful} cancellation of a ticket order is considered normal only when
(1) this ticket order exists in the database, 
(2) is in a valid state, and
(3) is associated with the logged-in user in session information.
Without comprehensive logging and querying of relevant information in the database and API execution context (session environmental information),
it is not informative to discriminate such a log event.
This brings the challenge of \textbf{C1: Observability Beyond Logs}, which means that the true normalities can be defined beyond the instrumented logs (e.g., database states and session environmental information).

However, comprehensive logging can introduce over detailed (or noisy) logs,
which contributes to learning false positive rules (see a real-world example in Section~\ref{sec:motivating-example}),
and it also incurs significant runtime overhead.
This brings the challenge of \textbf{C2: True Invariants from Noisy Log Instances}, which means that learning models or rules from noisy and lengthy log instances is challenging.

In this work, we propose \techname (\textbf{M}ining \textbf{I}nvariants for a\textbf{N}omaly d\textbf{E}tection via \textbf{S}chema),
which infers explainable API invariants for anomaly detection from the \textit{schema level}
instead of from detailed raw log instances.
\techname can
(1) infer more causal normalities and 
(2) capture crucial information beyond the log instrumentation.
Our rationale is that API specifications partially define the application's normalities.
Thus, specification-derived rules can deductively report anomalies in an explainable manner.
Compared to inferring specifications from noisy log instances,
we use abstract and precise meta-information in the application, such as API signatures and database schema, to
derive the implicit specification,
allowing us to
\begin{itemize}[leftmargin=*,topsep=2pt]
  \item (1) explore multi-domain information (e.g., database state and session environmental information) and combining them with logs (to address \textbf{C1}) and,
  \item (2) exploit the most crucial information from abstract log schemas rather than concrete raw logs (to address \textbf{C2}).
\end{itemize}
To this end, \techname reduces the anomaly detection problem to
an API specification inference problem.
Based on the inferred invariants, \techname can detect anomalies by checking the consistency between the logs and the inferred invariants.

Specifically, \techname infers implicit log invariants among APIs, database, and environmental information,
e.g.,
(1) ``\textit{what is the legitimate database state or environmental information required to call an API?}''
or
(2) ``\textit{what are the constraints that must be satisfied between different API calls?}''.
To achieve this, we build an augmented ER (entity-relationship) diagram
to capture the API-DB constraints, API-API constraints, and API-Env constraints
in the subject web application.
By normalizing API signatures into entities,
we can combine all the new and original entities and
use LLMs to infer their constraints, such as reference constraints (e.g., foreign keys), not-null constraints, equality constraints, etc.
Specifically, we convert each API (e.g., \texttt{cancelOrder(orderId, loginId)}) into a entity type (\texttt{cancelOrder}) with a set of attributes (\texttt{orderId} and \texttt{loginId}).
Then, we augment the original ER diagram in the database to an extended ER diagram containing those API entity types, 
allowing \techname to infer their implicit constraints as invariants using LLMs.
As a result, we can infer a specification such as
\begin{center}
\fbox{
\begin{minipage}{0.45\textwidth} 
For each entity \texttt{cancelOrder} with an attribute \texttt{orderId} (derived from the API, referred to as \textit{cancelOrder.orderId}),
there \textit{must} exist an entity \texttt{order},
such that \texttt{cancelOrder.orderId = order.id} and
\texttt{order.status = \texttt{"}paid\texttt{"}}.
\end{minipage}
}
\end{center}
This inferred specification indicates that 
an API call of \texttt{cancel\-Order} in the log 
(1) must have a corresponding record in the database table \textbf{order} with the corresponding order id and
(2) the corresponding order record must be in \textit{``paid''} status.
By inferring the invariants at the schema level, we can
(1) avoid inducing false positive invariants caused by accidental noisy correlations between two instances and
(2) build invariants connecting logged APIs with the internal database state.
In addition, we refine and filter the invariants by testing them
against a collection of known normal logs and retaining those that do not generate false alarms.
Finally, each invariant is translated into executable Python code
for runtime log verification.

We extensively evaluate \techname on web-tamper attacks on the
benchmarks of \trainticket and \nicefish against baselines such as LogRobust~\cite{zhang2019robust}, LogFormer~\cite{guo2024logformer}, and WebNorm~\cite{liao2024detecting}.
The results show that \techname achieves high recall (more than 14\% over LogRobust, LogFormer, and WebNorm) with almost zero false positive in detecting anomalies. Also, evaluation on three extra popular web applications, \gitea~\cite{gitea}, \mastodon~\cite{mastodon}, and \nextcloud~\cite{nextcloud}, shows that \techname can generalize well to popular industrial web applications.
Additionally, the performance of generating invariants is stable across a variety of LLMs,
indicating a new state-of-the-art anomaly detector for operating web applications.

In summary, the contributions of this paper are as follows:
\begin{itemize}[leftmargin=*,topsep=2pt]
    \item \textbf{Methodology.} We propose \techname, a schema-based specification-mining technique that unifies API signatures, database schema, and session environmental information, allowing us to exploit the most crucial features in abundant log structures and explore database states not instrumented in logs.
    \item \textbf{Tool.} We implement \techname as a framework which can be applied to any Java-based web application with available database schemas, facilitating real-world deployment.
    \item \textbf{Benchmark.}
        We build a web-tamper attack dataset, consisting of 31 types of attacks, that can successfully compromise
        known open-source web applications such as \trainticket and \nicefish.
        Thus, attacks are dynamic and replicable, allowing users to reproduce them with regenerated abnormal logs.
    \item \textbf{Evaluation.}
        We extensively evaluate \techname on the benchmark against state-of-the-art anomaly detectors such as LogRobust, LogFormer, and WebNorm, demonstrating its effectiveness in detecting web attacks and establishing a new state-of-the-art.
\end{itemize}

Given the space limit, more demos, source code, and experimental results are available at \url{https://sites.google.com/view/mines-anomaly-detection/home} \cite{mines-website}.

%% file: chaps/02-motivation-example.tex
\section{Motivating Example}\label{sec:motivating-example}

\input{floats/2-motivation-example}

\autoref{fig:motivation-example} shows abnormal logs caused by a web attack on the \trainticket system \cite{trainticketsystem},
which allows an attacker to successfully refund an order twice.
In this real-world example, the normal logs for refunding an order look very similar to the attack-incurred abnormal logs,
which causes state-of-the-art machine-learning based solutions such as LogFormer~\cite{guo2024logformer} and LogRobust~\cite{zhang2019robust} to fail to report the alarm.
In addition, the rule-learning based solution \cite{liao2024detecting} is misled to learn superficial rules (i.e., by capturing incorrect factors) from the normal logs,
leading to false negatives.

\noindent\textbf{Normal Logs and Their Semantics.}
The blue dashed rectangle in \autoref{fig:motivation-example} shows normal logs of refunding a ticket in the \trainticket system.
For clarity, we simplify the logs containing two API calls:
\begin{itemize}[leftmargin=*,topsep=2pt]
	\item \textbf{API of queryOrder}:
        The log of this API indicates a query for existing orders.
        Users usually call this API so that they can choose a specific order to refund in the frontend.
	\item \textbf{API of refundOrder}:
        This log indicates canceling an order and refunding the money back to a user in the backend.
\end{itemize}
In normal scenarios,
a user must query the orders before selecting one to refund.
As a result, the logs of these two APIs can repetitively occur
and form a pattern due to the design.
However, such a normality of correlation
does not indicate causality (i.e., the true normalities of calling \textbf{refundOrder}),
which misleads existing solutions to learn false predictions or summarize incorrect rules.

\noindent\textbf{Abnormal Logs Caused by Attacks.}
Unfortunately,
based on vulnerabilities in the exposed APIs in \trainticket,
a abnormal user can receive a refund multiple times.
Such abnormal logs from the attack are shown in the red dashed rectangle in \autoref{fig:motivation-example}.
Specifically, the attacker can call the exposed \textbf{refundOrder} an additional time to receive the extra refund.
We tried model-learning based solutions such as LogFormer \cite{guo2024logformer} and LogRobust \cite{zhang2019robust}
and rule-learning based solutions such as WebNorm \cite{liao2024detecting} to detect the anomaly,
finding their false negatives in practice as follows:

\begin{itemize}[leftmargin=*,topsep=2pt]
	\item \textbf{Model-learning Based Solutions.}
      Solutions such as LogRobust~\cite{zhang2019robust} and LogFormer~\cite{guo2024logformer} suffer from the
      \textit{unexpected} subtle differences between normal logs (in the blue dashed box) and abnormal logs (in the red dashed box).
      Generally, these approaches project logs into an embedding space for their predictions \cite{du2017deeplog}.
      Due to the textual similarity between normal and abnormal logs,
      the models have high confidence in reporting abnormal logs as normal.
\item \textbf{Rule-learning Based Solutions.}
      In contrast, WebNorm summarizes a superficial invariant as first-order logic from the log instances,
      indicating that
      (1) there must be a log of \textbf{queryOrder} occurring before a log of \textbf{refundOrder}
      and
      (2) the value of \texttt{orderId} in the log of \textbf{refundOrder} must be equal to the value of id in the log of \textbf{queryOrder}.
      This invariant (a.k.a., detection rule) is superficial because
      the normality (or specification) of the log of \textbf{refundOrder} actually depends on some database states instead of the appearance of the log of \textbf{queryOrder}.
      Even worse, the attack-incurred abnormal logs well match the false invariant,
      causing a false negative.
\end{itemize}

As shown in the green rectangle in \autoref{fig:motivation-example},
the true normality depends on the database state of the \textbf{order} table.
Specifically, given an order to refund, its normality depends on
whether there is a corresponding order record in the \textbf{order} table
with its status as ``\textit{paid}'' (see two tables under \textit{``Information beyond Logs''} in \autoref{fig:motivation-example}).

\noindent\textbf{Technical Challenges.}
To detect the aforementioned log anomalies,
we need to address the following technical challenges:

\begin{itemize}[leftmargin=*,topsep=2pt]
    \item \textbf{C1: Observability Beyond Logs.}
        The true normalities (or invariants) can be defined beyond the instrumented logs.
        In this example, we need to connect the logs, database states, and even session information to define a true invariant.
        However, exhaustively instrumenting database states into logs can incur large overhead.
        Therefore, we need to address \textit{``how do we achieve the required operational observability beyond log instrumentation?''}.
        Moreover, the database states can be volatile and dynamic,
        so we also need to address how to synchronize its frequent changes with the ever-growing logs.
    \item \textbf{C2: True Invariants from Noisy Log Instances.}
        Learning models or rules from noisy and lengthy log instances is challenging.
        Inductive deep learning solutions can sometimes capture correlation instead of causality,
        leading to the well-known spurious correlation problem \cite{ye2024spurious}.
        We need to address ``\textit{how to capture the most crucial facts and features to define the log normalities?}''.
\end{itemize}

In this work, we propose \techname, which detects runtime anomalies by inferring explainable invariants,
as shown in the green box in \autoref{fig:motivation-example},
from a schema level.
In contrast to all the state-of-the-art methods that learn from log instances,
we learn normalities and invariants from API signatures and database schema
in the subject web application.
This meta-information is more abstract and precise,
allowing us to
(1) explore additional information (e.g., database state and session environmental information) even if it is not in the logs (to address \textbf{C1}).
(2) exploit the most crucial information from the log structures (to address \textbf{C2}).

%% file: floats/2-motivation-example.tex
\begin{figure*}
	\centering
	\includegraphics[width=\motivatingwidth]{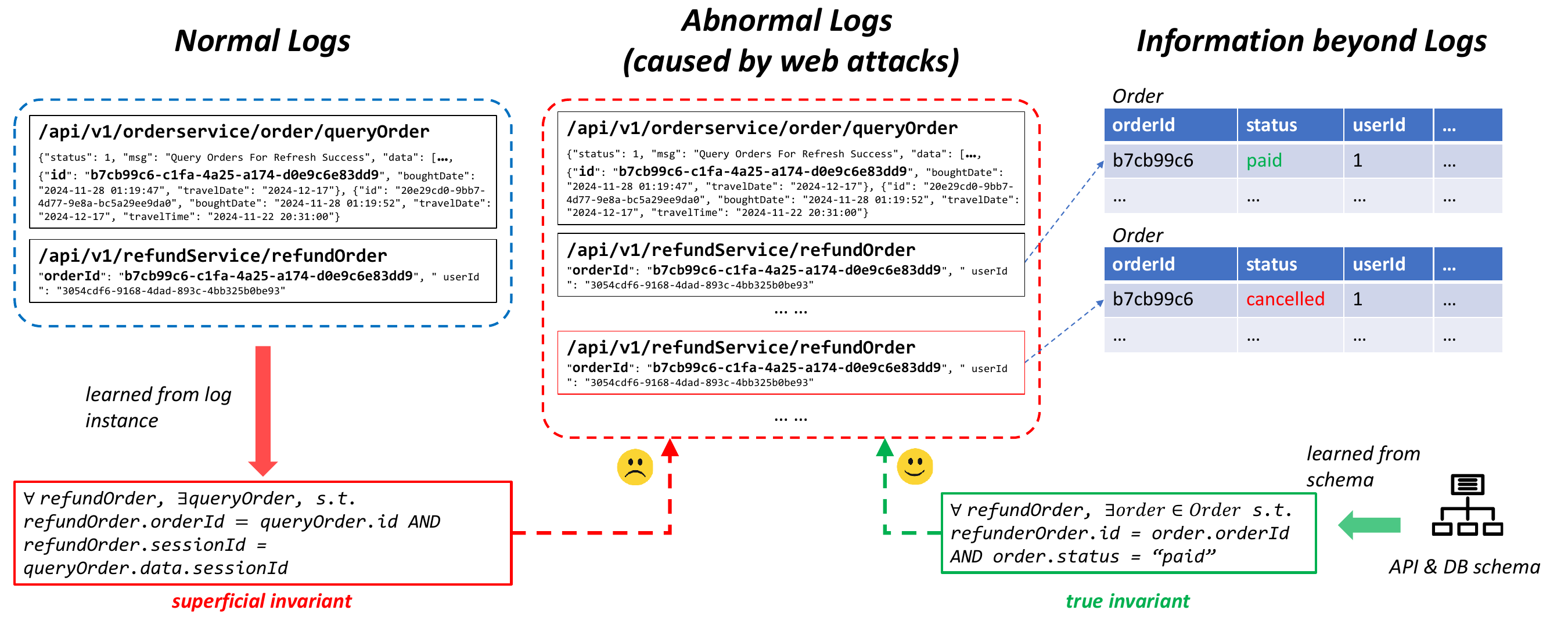}
	\caption{
    A log anomaly example caused by a real web attack on \trainticket system \cite{trainticketsystem}, 
    which can successfully refund an order twice.
    The attack logs are similar to the normal logs, which makes log classification/regression models such as LogRobust \cite{zhang2019robust} and LogFormer \cite{guo2024logformer} ineffective.
    In addition, rule-learning based solution such as WebNorm \cite{liao2024detecting} summarize a superficial rule from normal logs,
    which fails to detect such anomaly.
    }
	\Description{
This figure illustrates a real log anomaly example from the Train-Ticket web system. 
It compares normal and abnormal API logs to show why existing log-based anomaly detection fails.
On the left, normal logs show a sequence where a user first queries an order 
(`/api/v1/orderservice/order/queryOrder`) and then refunds it 
(`/api/v1/refundService/refundOrder`). 
A superficial invariant learned from these logs states that a refund must follow a query 
with the same orderId and sessionId. 
In the middle, abnormal logs caused by a web attack repeat the refundOrder API, 
allowing a user to receive multiple refunds even though the logs appear normal. 
On the right, database tables show that the order's status changes from ``paid'' to ``cancelled''. 
A schema-derived true invariant requires that a refundOrder must correspond to an order 
with status ``paid''. 
The figure highlights that rule-learning from raw logs (like WebNorm) learns 
superficial invariants, while schema-level reasoning (like MINES) captures the 
true causal invariant and can detect such anomalies.
    }
	\label{fig:motivation-example}
\end{figure*} 

%% file: chaps/03-method.tex
\section{Approach}

\input{floats/3-method}

\autoref{fig:method} shows an overview of \techname,
which parses the API signatures and database schema from the web application and
converts them into an augmented entity-relationship diagram (e.g., \autoref{fig:er-example}).
In such a diagram, entity types consist of not only database tables but also API signatures and environmental information (e.g., sessions).
Specifically, for each API signature,
we convert it into an entity type, where the name is the API name and the attributes are the API parameters and return values (e.g., \autoref{fig:api-example}).
For environmental information, we place it into a separate entity type.
Then, we reduce the anomaly detection problem into a constraint-inference problem for the ER diagram.
To this end, we apply a two-stage inference in our approach:
\begin{itemize}[leftmargin=*,topsep=2pt]
	\item \textbf{Stage One: Entity-Relationship Inference.}
	      At this stage, we infer relationships among entity types.
	      The inferred relationships allow us to join two entity types (i.e., tables).
	      For example, this allows us to join the \texttt{orderId} field in the \textbf{API:refundOrder} entity type to the \texttt{orderId} field in the \textbf{orders} entity type (\autoref{fig:er-example}).
	\item \textbf{Stage Two: Invariant Inference.}
	      At this stage, we join the tables by the inferred foreign keys from the first stage.
	      For each joined table (e.g., between \textbf{API:refundOrder} entity type and \textbf{orders} entity type), we further infer not-null constraints, equality constraints, check constraints, etc., on their attributes.
	      These constraints consist of both intra-entity constraints (e.g., \texttt{price} should be a positive number) and inter-entity constraints (e.g., \texttt{orderId} in \textbf{API:refundOrder} should match with \texttt{orderId} in \textbf{orders}).
	      For example, this allows us to infer that the \texttt{status} attribute needs to be \textit{``paid''} (\autoref{fig:inv-inference}).
	      Finally, the inferred constraints are translated into Python code as executable invariants.
\end{itemize}

Both stages require state-of-the-art LLMs (e.g., ChatGPT, Claude, and DeepSeek)
to infer the constraints regarding the semantics.
To mitigate their potential hallucinations,
we run the subject web application in a secure environment to obtain normal logs.
Any generated invariants that raise false alarms on the normal logs will be removed to reduce false rules.

\subsection{Information Requirements}

Web applications differ widely in their architectures and implementations. To ensure broad applicability, \techname requires only a minimal and commonly available set of information from the target web application:
\begin{itemize}[leftmargin=*,topsep=2pt]
    \item \textbf{API Signatures:} Definitions of API endpoints, parameter types, and return types.
    \item \textbf{Database Schema:} Table structures and attribute types.
    \item \textbf{Contextual Information:} API invocation context, including session structures and access methods.
    \item \textbf{API Logs:} Collected records of API requests.
    \item \textbf{Database Binary Logs:} Historical records of database changes.
\end{itemize}

\subsection{Schema Parsing}
\label{sec:specifications}

\begin{figure}
	\centering
	\subfloat[API Signature]{\label{fig:api-signature}\includegraphics[width=.35\columnwidth]{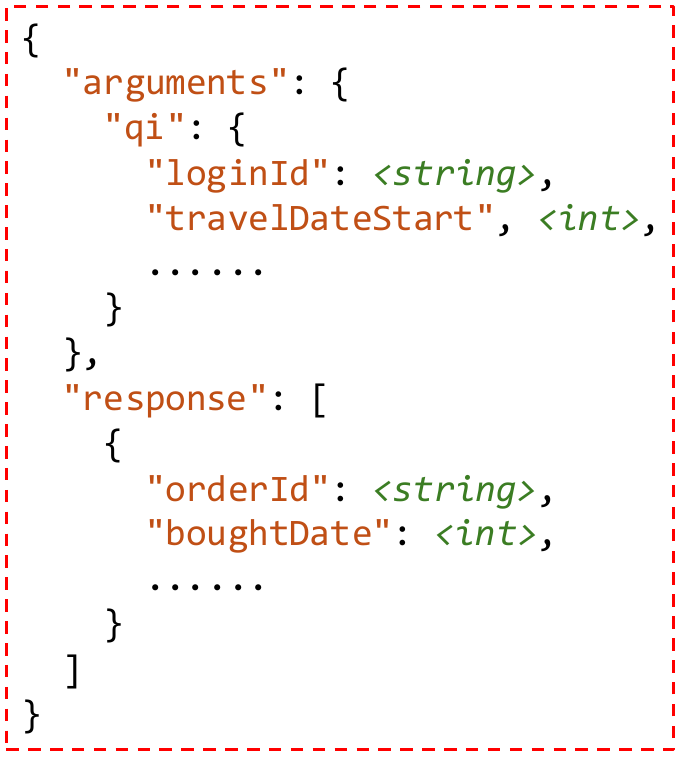}}
	\qquad
	\subfloat[API Entity Type]{\label{fig:api-entity}\includegraphics[width=0.26\columnwidth]{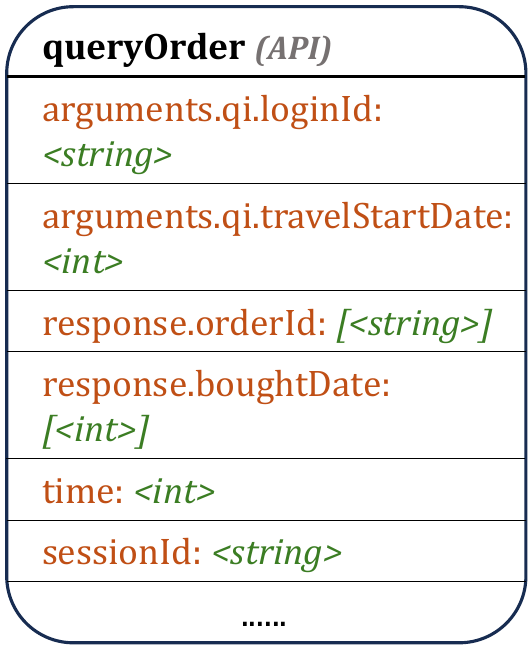}}
	\caption{Example of Converting an API Signature to an API Entity Type.}
	\label{fig:api-example}
	\Description{
This figure illustrates how MINES converts an API signature into an API entity type. 
Subfigure (a) shows a JSON-style API definition where the arguments and response fields 
include attributes such as loginId, travelDateStart, orderId, and boughtDate. 
Subfigure (b) shows the corresponding API entity type representation, in which each 
argument and response field is flattened into a tabular schema with data types like 
<string> and <int>, forming a structured entity called queryOrder with attributes such 
as arguments.qi.loginId, response.orderId, response.boughtDate, time, and sessionId.
	}
\end{figure}

We parse the subject web application into three types of meta-information for logs,
i.e., API signatures, database schema, and environmental information (e.g., sessions).
All are processed into the database schema for further analysis.

\noindent\textbf{Converting API Signatures to Schema.}
We convert each API signature into an \textit{action} entity type,
where the API name serves as the table name and the API parameters serve as the table attributes.
Intuitively, each API signature is mapped into an entity type, while each API log is mapped into an entity instance.
However, unlike traditional relational database table schemas where
each attribute is a primitive type (e.g., string, int, or blob),
an API can take complex objects (e.g., \texttt{OrderInfo} object) as input.
Therefore, we parse the object tree structure into a flattened parameter list
as shown in \autoref{fig:api-example}.
Specifically, given a complex input object as tree $\tau$ and a threshold $th$,
we expand $\tau$ into the list according to the depth of $th$.

\begin{figure}
	\centering
	\includegraphics[width=.25\columnwidth]{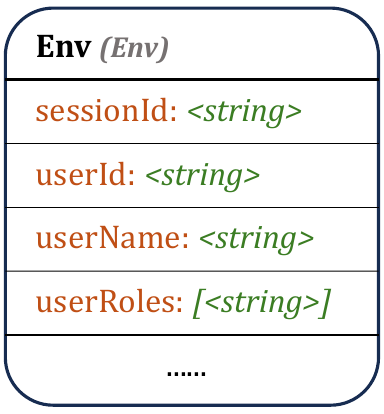}
	\caption{Example of Environmental Information.}
	\label{fig:env-example}
	\Description{This figure illustrates an example of environmental information used by MINES. 
It displays a structured Env entity containing session-related attributes such as 
sessionId, userId, userName, and userRoles. Each field is associated with its data type, 
illustrating how contextual or session information is represented in schema form for 
further invariant inference.}
\end{figure}

\noindent\textbf{Converting Environmental Information to Schema.}
Web applications utilize sessions for user authentication and authorization.
There are many different implementations of session management, such as authentication tokens, cookies, and session IDs.
For example, \autoref{fig:env-example} shows an example of environmental information in the \trainticket system.
The environmental information stores the current user ID number, current user role, and current user name.
By collecting and analyzing this information, we can ensure better security and a more personalized user experience.
To capture structured logs and extract environmental information, we implemented a new instrumentation framework in Java, which is designed to capture environmental information from API handling methods.

\subsection{Relationship Inference}
\label{sec:related-entity-prediction}

\input{floats/3-propmt-relation}

This step aims to build the relationships from API entity types to other entity types.
Specifically, \techname infers three types of relationships:
(1) API-DB Relationships, i.e., foreign keys from API entity types to database tables,
(2) API-API Relationships, i.e., temporal dependencies between different API entity types,
and (3) API-Env Relationships, i.e., relevance of environmental information for APIs.

Formally, we define the three relationships as follows:
\begin{itemize}[leftmargin=*,topsep=2pt]
	\item \textbf{API-DB Relationships.} Given an API $E_{API}$ with an attribute $a_{API}$ and a database table $E_{DB}$ with an attribute $a_{DB}$, for every API log instance $e_{API} \in E_{API}$, there exists a database row $e_{DB} \in E_{DB}$ such that $e_{API}.a_{API} = e_{DB}.a_{DB}$.
	\item \textbf{API-API Relationships.} Given a source API  $E_{API1}$ and a target API $E_{API2}$, for every API log instance $e_{API1} \in E_{API1}$, there exists an API instance $e_{API2} \in E_{API2}$ such that $0 < e_{API1}.time - e_{API2}.time < \delta$ and $e_{API1}.session = e_{API2}.session$.
	\item \textbf{API-Env Relationships.} Given an API $E_{API}$ and environmental information $E_{Env}$, for every API log instance $e_{API} \in E_{API}$, there exists an environmental entity instance $e_{Env} \in E_{Env}$ such that $e_{API}.session = e_{Env}.session$.
\end{itemize}

\techname infers the relationships (1) using a LLM, and then (2) refines the relationships using heuristic rules.
For each pair of entity types, \techname first asks the LLM to find all potential relationships.
\autoref{fig:prompt-relation} shows the prompt used for relationship inference.
The prompt includes a brief introduction to the definition of the task, the input/output format, and an in-context learning example.
Due to space limitations, the detailed instructions are available in the code repository \cite{mines-website}.

\techname filters out false relationships using the following heuristic rules:
\begin{itemize}[leftmargin=*,topsep=2pt]
    \item \textbf{API-DB Relationships:} Discarded if the attribute value overlap between columns is below a threshold.
    \item \textbf{API-API Relationships:} Discarded if their sequence probability, computed by a hidden Markov model (HMM) trained on API invocation logs, is below a threshold.
    \item \textbf{API-Env Relationships:} Discarded if the relevant environmental information is not present in the session logs.
\end{itemize}
The thresholds used in these heuristics are not specific to any application and can be easily adjusted for new projects with minimal effort, typically requiring only a small validation set or basic log statistics. 
Also, these heuristic rules serve only as a coarse filtering step; they do not play a central role in the overall analysis, but rather help to efficiently eliminate obvious false relationships before further processing.


\input{floats/3-er-example}

\autoref{fig:er-example} shows an example of inferred relationships.
The focal API entity type \textbf{refundOrder} has three relationships with other entity types,
i.e., (1) a foreign key relationship with the \textbf{orders} table on the \texttt{orderId} attribute,
(2) a temporal dependency with the \textbf{getOrder} API,
and (3) relevancy to the environmental information.

\subsection{Invariant Generation}
\label{sec:invariant-deduction}

\input{floats/3-inv-example}

\input{floats/3-refinement}

To generate invariants,
\techname first joins the entity types based on the inferred relationships.
Then, \techname generates candidate invariants on the joined tables.
After that, \techname refines the invariants using training logs.
\autoref{fig:inv-inference} shows the process of invariant generation.

\noindent\textbf{Joining Entities.} 
Given any two entity types $E_1$ and $E_2$ with a relationship (API-DB, API-API, or API-Env), we perform a left outer join on their corresponding tables to preserve all API log instances. The join strategy follows the relationship type:
\begin{itemize}[leftmargin=*,topsep=2pt]
    \item \textbf{API-DB Relationships:}  
    Correspond to foreign key constraints. For $E_{API}$ and $E_{DB}$ with relationship on $a_{API}$ and $a_{DB}$, the join is $E_{API} \times E_{DB}$, filtered by $a_{API} = a_{DB}$.
    \item \textbf{API-API Relationships:}  
    Represent temporal dependencies. The join is $E_{API1} \times E_{API2}$, filtered by the temporal condition $0 < e_{API1}.\mathit{time} - e_{API2}.\mathit{time} < \delta$.
    \item \textbf{API-Env Relationships:}  
    Based on environmental context, typically joined via session ID, analogous to a foreign key.
\end{itemize}
The resulting joined table (e.g., \textbf{API:refundOrder-orders-Env} in \autoref{fig:inv-inference}) contains extra information beyond original API logs.

\noindent\textbf{Candidate Invariant Generation.} In this step, given the joined table structure, \techname asks LLMs to generate candidate invariants as Python code that checks the constraints on the joined table. \autoref{fig:inv-inference} shows an example of candidate invariant generation. The input is the joined table structure, which include table names, attributes, and their types. The output is pieces of Python code that checks the constraints on the joined table.

In the prompt, we provide a brief introduction to the task, the input/output format, and an in-context learning example. Due to space limitations, the detailed instructions are available in the code repository \cite{mines-website}.

To help LLMs to generate invariants we want, the prompt guide the LLMs to generate invariants in five categories:
\begin{enumerate}[leftmargin=*,topsep=2pt]
	\item \textbf{Common-sense constraints.} Constraints that are generally applicable to the request (e.g., \texttt{price} should be a positive number).
	\item \textbf{Format constraints.} Constraints on the format or type of the field (e.g., valid \texttt{email} format).
	\item \textbf{Database constraints.} Constraints between the API arguments and the database entity (e.g., \texttt{orderId} should match the order entity).
	\item \textbf{Environment constraints.} Constraints between the API arguments and the environmental information (e.g., \texttt{userId} should match the user ID in the session).
	\item \textbf{Related API constraints.} Constraints between the request arguments and the responses of related APIs (e.g., data flow between different APIs).
\end{enumerate}

\noindent\textbf{Invariant Refinement.} \techname iteratively refines the candidate invariants using training logs. Each candidate invariant is evaluated against the training logs. If any violations are detected, i.e., the invariant fails to hold on normal logs, these violations, along with relevant contextual information and error messages, are fed back to the LLM within the same conversation thread. The LLM is then prompted to revise or discard the problematic invariant. This creates a feedback loop in which the LLM iteratively refines its output based on concrete examples of failure. To prevent indefinite retries, this loop is bounded: if the LLM fails to produce a valid invariant within a fixed number of iterations, the system discards the candidate altogether by returning an empty result.

\subsection{Runtime Verification}
\label{sec:binary-log}

\noindent\textbf{Offline Attack Detection.} Based on the generated invariants, we run them against the logs and database records.
Our method works in an offline manner,
which means that the evaluation of the invariants will be carried out after a period of time following API execution.
Offline attack detection minimizes online overhead.

In the offline attack detection setting, once the invariants are learned, the system operates only with two main components: a log collector and an offline checker.
\begin{itemize}[leftmargin=*,topsep=2pt]
	\item The \textbf{log collector} continuously collects logs during runtime as a producer.
	\item The \textbf{offline checker} executes the learned invariants (compiled as Python code) against the incoming logs to detect violations. Multiple checkers can be deployed in parallel to improve throughput and reduce latency.
\end{itemize}

\noindent\textbf{Binary Log History Tracking.} In the offline setting, we need to address the challenge of database dynamics,
i.e., how do we track the historical database records even if the database state changes.


To this end, we replay the binary log, which records all changes to database records and is typically enabled by default in modern web applications, to restore the database state during invariant evaluation. By extracting and applying binary log records, we ensure consistent and accurate evaluation of invariants, mitigating issues from stateful data changes.

%% file: floats/3-method.tex
\begin{figure*}
	\centering
	\includegraphics[width=.7\textwidth]{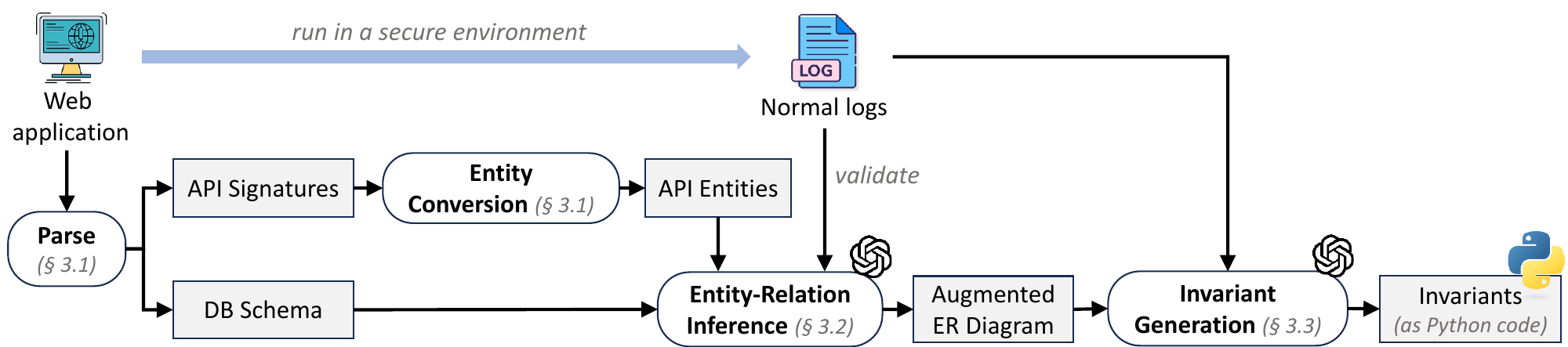}
	\caption{Approach Overview: Given a web application, \techname parses API signatures and database schema into an augmented ER (entity-relation) diagram. By inferring the reference constraints and customized constraints over the generated diagram by LLM, we infer the invariants as Python code for runtime verification. In addition, to avoid hallucination of LLM, we use normal logs to refine the generated constraints.}
	\Description{
		This figure presents the overall workflow of MINES for explainable anomaly detection.  
Starting from a web application, MINES parses both API signatures and database schema.  
The API signatures are converted into API entities, and together with the database schema,  
they are passed to the Entity-Relation Inference stage, where large language models infer  
potential reference and customized constraints between entities.  
The resulting augmented entity-relationship diagram is then used for invariant generation,  
where LLMs synthesize executable Python invariants that describe valid relationships and  
conditions among API calls, database states, and session information.  
Normal logs are used to validate and refine these invariants to remove false or spurious  
rules. The final output is a set of Python invariants that can be used for runtime anomaly  
verification.  
The figure also shows that the process runs in a secure environment and highlights three  
main stages: parsing, entity-relation inference, and invariant generation.  
	}
	\label{fig:method}
\end{figure*} 

%% file: floats/3-propmt-relation.tex
\begin{figure}
	\centering
	\begin{framed}
		\raggedright
		\footnotesize
		\linespread{1.0}

		\textbf{\# Identity}

		You are a software engineer who is extremely good at understanding business logic and user requirements for web applications.
		Given two entities, your task is to find out if there are any relationships between them.

		\textcolor{gray}{\textit{[relationship definition] [input/output format] [in-context learning example]}}

		\textbf{\# Input}

		\definecolor{mygreen}{RGB}{150, 255, 150}

		- Focal Entity Type: \colorbox{mygreen}{\textbf{\path{cancel.service.CancelServiceImpl.cancelOrder} }} \colorbox{mygreen}{ \texttt{\{ "userId": <str value>, "orderId": <str value> \}}}

		- Target Entity Type: \colorbox{mygreen}{\textbf{\path{users}} \texttt{\{ "id": <str value>, "name": <str value> \}}}

		\textbf{\#\# Output:}

		\colorbox{pink}{

		\parbox{.95\columnwidth}{

		\texttt{<thought>} \textcolor{gray}{\textit{[chain of thought]}} \texttt{</thought>}

		{

		\ttfamily

		\textasciigrave \textasciigrave \textasciigrave json


		\{

		\ \ "relationships": [

		\ \ \ \ "from\_column": "userId",

		\ \ \ \ "to\_column": "id",

		\ \ ]

		\}

		\textquotesingle \textquotesingle \textquotesingle

		}

		}

		}

	\end{framed}
	\caption{Prompt used for relationship inference. The green boxes represent the input information, and the pink box represents the output of the LLM.}
	\label{fig:prompt-relation}
	\Description{Prompt used for invariant relationship inference.}
\end{figure}

%% file: floats/3-er-example.tex
\begin{figure}
	\centering
	\includegraphics[width=.5\columnwidth]{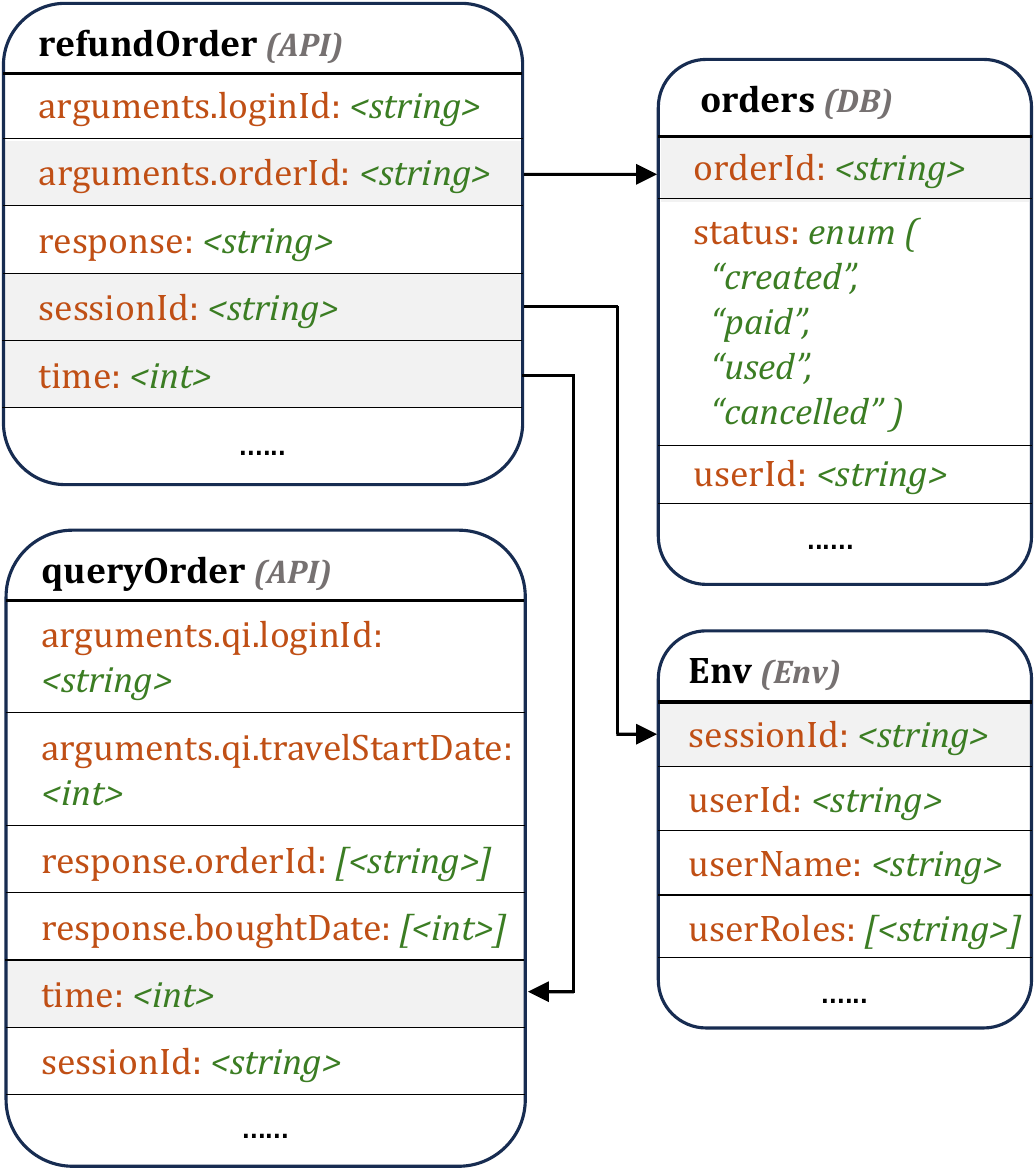}
	\caption{Example of an augmented entity-relation diagram and inferred relationships. The API \textbf{refundOrder} has three relationships with the database table \textbf{orders}, another API \textbf{queryOrder}, and environmental information \textbf{Env}.}
	\label{fig:er-example}
	\Description{
This figure illustrates an augmented entity-relationship diagram showing inferred relationships among 
API entities, database tables, and environmental information. The API refundOrder is linked to three 
entities: the database table orders, another API entity queryOrder, and the environmental entity Env. 
The diagram highlights different relationship types, foreign key links to the database, temporal or 
data dependencies between APIs, and contextual associations with the environment. Each entity lists 
its attributes, such as orderId, userId, and status, with corresponding data types and enumerated values.
	}
\end{figure}

%% file: floats/3-inv-example.tex
\begin{figure}
	\centering
	\includegraphics[width=.75\columnwidth]{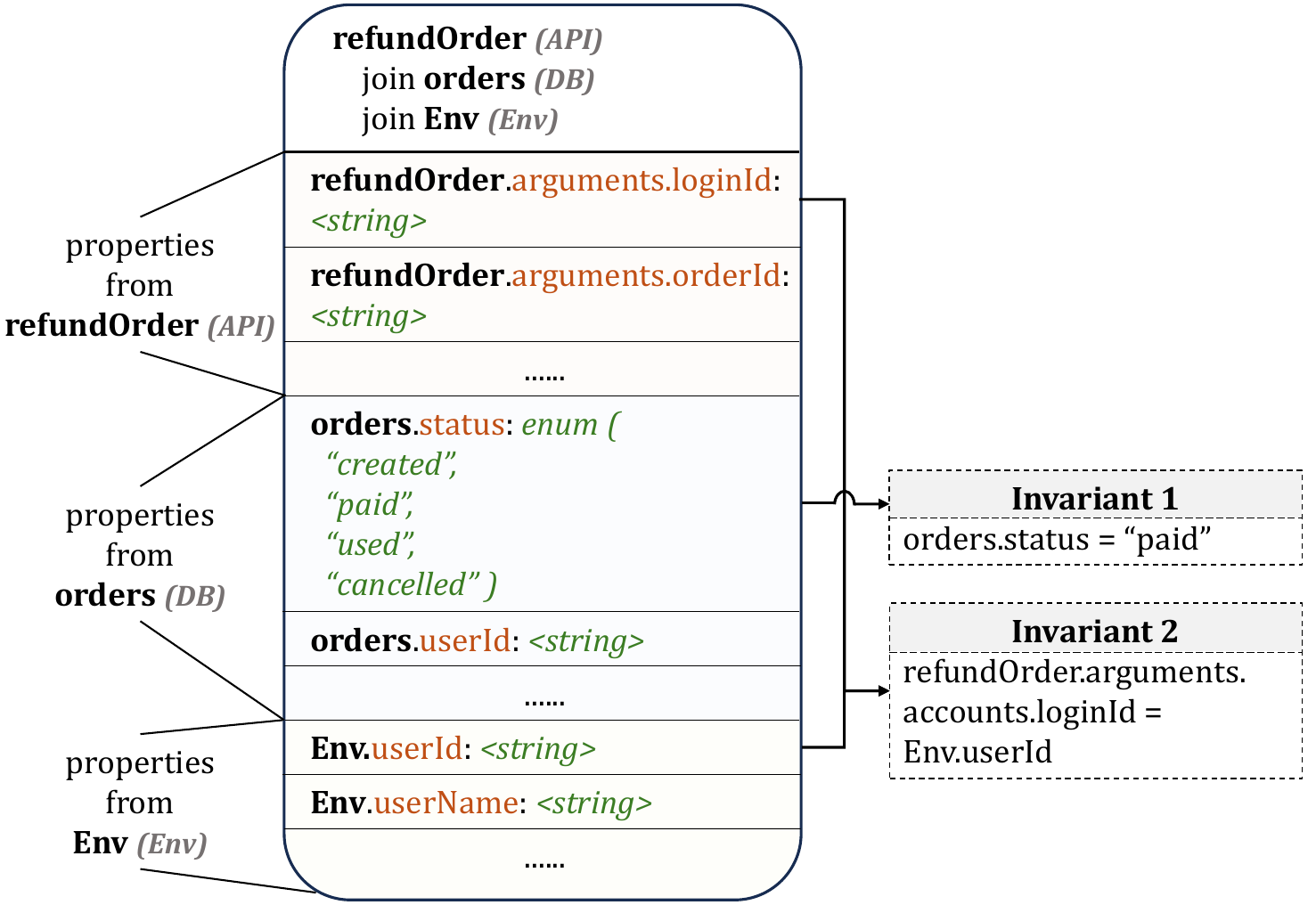}
	\caption{Candidate invariant generation over the joined table. The whole table consists of attributes from three tables, including API refundOrder, database table orders, and environmental information Env. Invariants are inferred from the joined table.}
	\label{fig:inv-inference}
	\Description{
This figure shows the process of generating candidate invariants from the joined tables that combine 
attributes from API refundOrder, database table orders, and environmental information Env. 
The figure illustrates how properties from each source are merged into a unified schema to infer 
constraints such as orders.status equals "paid" and refundOrder.arguments.loginId equals Env.userId. 
These inferred invariants capture logical relationships between API inputs, database states, 
and user session information for anomaly detection.

	}
\end{figure}

%% file: floats/3-refinement.tex
\begin{figure}
	\centering
	\includegraphics[width=.7\columnwidth]{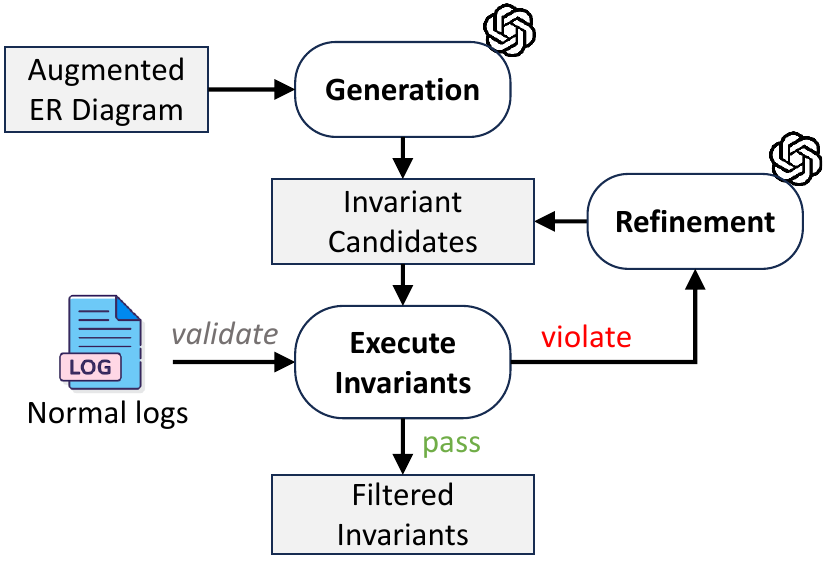}
	\caption{Invariant generation process. \techname first generates invariant candidates from the augmented ER diagram. Then it iteratively refines the candidates by executing the invariants against the training normal logs. If any violations occur, the invariants are refined by feeding back to the LLM again with the error message.}
	\label{fig:refinement}
	\Description{
This figure depicts the invariant generation and refinement workflow in MINES. 
Starting from the augmented ER diagram, candidate invariants are generated by a large language model. 
These invariants are then validated against normal logs—if an invariant passes all checks, 
it becomes part of the filtered invariants; if violations occur, the system feeds the error 
back to the LLM for refinement. The loop continues until consistent and validated invariants 
are produced, ensuring both precision and robustness in anomaly detection.
		}
\end{figure}

%% file: chaps/04-experiments.tex
\section{Benchmark Construction}

Executing \techname requires not only API logs, but also database information, environmental information, and binary logs.
However, existing datasets for web applications only contain logs and do not provide necessary information for re-execution.
Therefore, we constructed two new datasets for two web applications.
These datasets provide dynamic and replicable scripts to generate both normal and attack logs, allowing us to reproduce logs or binary logs as needed.

To align with our baseline WebNorm~\cite{liao2024detecting}, we collected our datasets from the same two web applications: \trainticket~\cite{trainticketsystem} and \nicefish~\cite{nicefish}, which are widely used in the web development community. Our new datasets contain 31 types of attacks and cover 27 APIs.

\subsection{Dataset Construction}

To ensure dataset diversity, we generate both normal and abnormal logs using manually written scripts and LLM-based methods. The logs are categorized into four types:

\begin{itemize}[leftmargin=*,topsep=2pt]
	\item \textbf{\logsn}: Manually scripted normal logs, representing typical user operations organized as finite state machines.
	\item \textbf{\logln}: LLM-generated normal logs, using scripts from WebNorm~\cite{liao2024detecting} to predict actions from web application screenshots and task descriptions.
	\item \textbf{\logsa}: Manually scripted abnormal logs, targeting identified web abnormal endpoints by scripting abnormal scenarios for each application functionality.
	\item \textbf{\logia}: Abnormal-injected logs, created by injecting abnormal fields into normal logs from \logln, following the protocol in WebNorm~\cite{liao2024detecting}.
\end{itemize}

\input{floats/4-number}

\autoref{tab:script-statistics} and \autoref{tab:dataset-statistics} present the statistics of normal and abnormal scripts and tests for the \trainticket and \nicefish datasets.
	LLM-generated normal logs more closely reflect real user behavior, while manual scripts provide comprehensive coverage. For abnormal logs, both manual and injection-based methods are used to capture a wide range of attack scenarios. This ensures our dataset thoroughly encompasses all abnormal cases in WebNorm~\cite{liao2024detecting}.
	We constructed all identifiable normal and abnormal scenarios for both web applications. These scripts not only cover every normal and abnormal scenario in WebNorm~\cite{liao2024detecting}, but also increase diversity by targeting more APIs and abnormal cases.
	Detailed examples can be found at~\cite{mines-website}.

\subsection{Division of Training and Evaluation Datasets}

Given the four types of generated logs, we construct training and evaluation datasets separately for \techname and baselines.

	Rule-learning approaches, such as WebNorm~\cite{liao2024detecting} and our proposed \techname, require only normal logs for training. For this purpose, we use logs generated by \logsn, which comprehensively cover a wide range of normal operations. For evaluation, we employ logs generated by \logln, \logsa, and \logia to assess generalization and robustness.

	In contrast, model-learning approaches such as LogRobust~\cite{zhang2019robust} and LogFormer~\cite{guo2024logformer} rely on both normal and abnormal logs for training. Therefore, in addition to \logsn logs, their training sets also include normal logs generated by \logln and attack logs.


	It is important to note that model-learning baselines are trained with a larger volume of data than \techname. This design choice does not compromise fairness; instead, it favors the baselines by giving them access to more expressive training signals. Despite this advantage, \techname still achieves superior performance, demonstrating its effectiveness under a more constrained training setup.

\section{Experiments}

We evaluate our approach with the following research questions:
\begin{itemize}[leftmargin=*,topsep=2pt]
	\item \textbf{RQ1:} What is the overall effectiveness and efficiency of our approach compared with the baselines?
	\item \textbf{RQ2:} How do different components (\textit{schema-based deduction}, \textit{contextual relationships}, \textit{binary log history tracking}) and the \textit{invariant refinement} process contribute to the performance and robustness of our approach?
	\item \textbf{RQ3:} How robust is our approach when equipped with different language models?
	\item \textbf{RQ4:} How does the quality and consistency of system naming conventions affect the performance of our approach?
	\item \textbf{RQ5:} Can our approach generalize to popular real-world web applications?
\end{itemize}

\subsection{Setup}

\noindent\textbf{Baselines.}
We compare \techname with two categories of baselines:
\begin{itemize}[leftmargin=*,topsep=2pt]
	\item \textbf{Model-learning based approaches:} We chose LogRobust~\cite{zhang2019robust} and LogFormer~\cite{guo2024logformer} as baselines because they are the latest model-learning based applications and perform best among similar models, as shown in \cite{liao2024detecting,guo2024logformer}.
	\item \textbf{Rule-learning based approaches:} We compare \techname with WebNorm~\cite{liao2024detecting}, which is the only existing interpretable normality learning method.
\end{itemize}

\noindent\textbf{Benchmarks.}
We utilize the two benchmarks constructed in the previous section, \trainticket and \nicefish.

\noindent\textbf{Metrics.}
We use precision and recall as metrics.
For normal logs, we split logs into windows of 20 logs each.
A window is marked \textit{False Positive} (FP) if any attacks are detected;
otherwise, it is \textit{True Negative} (TN).
For attack logs, detection of any attacks results in \textit{True Positive} (TP) for all logs;
otherwise, they are \textit{False Negative} (FN).
Precision and recall are calculated as:
$\text{Precision} = \frac{TP}{TP + FP},$ and
$\text{Recall} = \frac{TP}{TP + FN}.$

\noindent\textbf{LLMs Used.}
We primarily use the GPT-4o model~\cite{gpt4o},
specifically version gpt-4o-2024-08-06,
for our evaluation.
To assess \techname's performance across different LLMs,
we also employ GPT-4o-mini (gpt-4o-mini-2024-07-18),
Claude 3.7 Sonnet~\cite{claude37} (claude-3-7-sonnet-20250219),
and DeepSeek-V3~\cite{deepseekv3} (deepseek-v3-241226).

\subsection{Overall Effectiveness and Efficiency (RQ1)}

\input{floats/4-overall}

We comprehensively evaluate the effectiveness and efficiency of \techname
against baselines on two datasets, as shown in \autoref{tab:overall-eval}.
\techname consistently outperforms baselines on both \trainticket and \nicefish.
Both \techname and WebNorm achieve 100\% precision due to refined invariants, ensuring accuracy.
In recall, \techname achieves 94.8\% on \trainticket and 91.7\% on \nicefish,
exceeding baselines by over 15\%,
demonstrating superior attack detection\footnote{The precision and recall values reported for WebNorm differ from those in the original WebNorm paper due to differences in experimental settings. Specifically, our evaluation covers a broader set of scenarios, which, combined with the invariant refinement process, leads to higher precision but lower recall compared to the original results.}.

\input{floats/4-two-types}

The attack data includes \logsa and \logia.
We evaluated \techname on these attacks separately,
as shown in \autoref{tab:two-type-attacks}.
\techname achieves a recall of 1.0 on \logia,
due to the easier detection of simulation attacks,
and nearly 0.9 on \logsa,
demonstrating its effectiveness on real attacks.

\input{floats/4-timecost.tex}

We further measure the training and evaluation performance of \techname, as shown in \autoref{tab:performance-eval}.
The training overhead and cost are acceptable,
and \techname achieves high evaluation throughput, $4\times 10^5$ logs per second on \trainticket and $2\times 10^5$ logs per second on \nicefish,
significantly faster than model-learning baselines.

\finding{\textbf{RQ1}: \techname achieves both high effectiveness and efficiency.
It outperforms baselines in attack detection with 100\% precision and over 15\% higher recall,
while maintaining acceptable training overhead and processing more than $2\times 10^5$ logs per second during evaluation.}

\subsection{Component Contribution (RQ2)}

\input{floats/4-ablation}

To understand how different components contribute to \techname's effectiveness and efficiency,
we conducted a comprehensive ablation study and analysis on prompt construction,
input representation, and invariant refinement.

\paragraph{Impact of Core Components.}
We first examine how individual components affect recall while maintaining 100\% precision,
as shown in \autoref{tab:ablation-study}.
Removing key elements such as API-DB, API-API, or API-Env relationships
causes the recall to drop from 0.948 to 0.820, 0.908, and 0.780, respectively.
Among them, API-Env contributes the most, followed by API-DB.
This demonstrates that contextual and structured information, linking API calls with their data and environment, is crucial for accurate invariant reasoning.
Similarly, omitting binary log history tracking, which aligns each API invocation with its historical database state,
significantly reduces recall, confirming the importance of temporal consistency.
Notably, WebNorm can be viewed as a composite ablation combining several of these removals,
explaining its lower recall in prior comparisons.

\paragraph{Effect of Deducing from Schemas.}
\input{floats/4-input-tokens}
\input{floats/4-hist-tokens}

We further compare \textit{deducing from schemas} with \textit{inducing from raw logs}.
Replacing schema-based prompts with raw logs decreases recall from 0.948 to 0.896,
showing that schemas provide cleaner and more abstract representations that generalize better than noisy instances.
In addition, schema-based deduction dramatically reduces token counts in prompts,
by up to two orders of magnitude across mean, geometric mean, and median values,
as shown in \autoref{tab:comparing-input-token-numbers} and illustrated in \autoref{fig:hist-tokens}.
This compression leads to lower inference cost and faster evaluation without loss of accuracy.

\paragraph{Effect of Invariant Refinement.}
\input{floats/5-refinement.tex}

\techname uses a two-stage pipeline for invariant generation:
(1) extracting candidate invariants from schemas and
(2) refining them with training logs.
The refinement step is crucial for filtering false positives,
reducing their rate from several percent to zero (\autoref{tab:inv-refinement}).
Without refinement, the system would produce excessive false alarms,
especially in large-scale deployments.
Refinement also mitigates LLM hallucinations through iterative feedback,
where invalid or overfitted invariants are revised or discarded based on validation logs.
Only a moderate amount of normal log data is required,
making this process lightweight yet essential for practical precision.

\finding{\textbf{RQ2:}
Each component of \techname, including schema-based deduction,
API-DB/API-Env/API-API relationships, binary log tracking, and invariant refinement,
plays a vital role in enhancing both recall and efficiency.
Schema-based inputs reduce token length by up to two orders of magnitude,
while refinement eliminates false positives and ensures robust, deployable precision.}

\subsection{Comparing LLMs (RQ3)}

\input{floats/4-different-llms}

We evaluated \techname using different LLMs, GPT-4o-mini, Claude 3.7, and DeepSeek-V3, on the \trainticket and \nicefish dataset to assess its robustness across models. As shown in \autoref{tab:comparing-llms}, Claude 3.7 and DeepSeek-V3 achieve comparable results to GPT-4o, demonstrating that \techname maintains consistent performance across different architectures.

GPT-4o-mini, however, performs slightly worse. To better understand this gap, we conducted a manual analysis of 10 representative cases where GPT-4o succeeded but GPT-4o-mini failed. In 9 out of 10 cases, GPT-4o-mini correctly described the invariants in natural language but failed to produce valid executable code, mainly due to formatting issues or semantic inconsistencies. This suggests that the performance gap is mainly attributable to limitations in code generation.

\finding{\textbf{RQ3}: \techname achieves consistent performance across LLMs, though smaller models like GPT-4o-mini may struggle with code generation despite adequate reasoning.}
\subsection{Effect of Naming Conventions (RQ4)}

\input{floats/5-naming.tex}

Large language models play a central role in \techname, as they interpret entity and attribute names to induce semantic relationships.
Consequently, \techname's effectiveness inherently depends on the clarity and consistency of naming conventions in the target system.
In our default setup, we assume that system identifiers (e.g., API names, table names, and field names) are reasonably descriptive, following common engineering practice.

To systematically assess the impact of naming quality, we conducted controlled experiments by modifying the original entity names using several strategies:
(1) replacing words with partial or extreme abbreviations,
and (2) applying stylistic variations such as snake case and camel case.
As shown in \autoref{tab:naming-conventions}, \techname remains robust under most conventional naming styles and moderate abbreviations.
However, its performance declines significantly when names are heavily abbreviated (e.g., replaced by single letters or meaningless tokens).
In such cases, the model mainly detects superficial format violations rather than deeper inter-attribute inconsistencies.

This result highlights that while \techname can tolerate moderate variation in naming,
it fundamentally relies on semantically meaningful identifiers to establish relationships across APIs, database fields, and environmental contexts.
Fortunately, such descriptive naming is widely adopted in real-world software systems,
so the dependency is realistic and manageable in practice.

\finding{\textbf{RQ4:}
\techname is generally robust to common naming styles and moderate abbreviations,
but its performance degrades when semantic clarity is lost.
This demonstrates that the approach relies on meaningful system naming to guide LLM reasoning.}

	\subsection{Generalization to Popular Web Applications (RQ5)}

	\input{floats/4-more-benchmarks.tex}
	\input{floats/4-more-benchmarks-res.tex}

	To evaluate \techname's generalization to popular web applications, we collected three GitHub repositories of web applications.
	We referred to Gitstar Ranking to select the most starred web applications. We filtered out applications that are not productive or learning projects.
	We got the first three applications: NextCloud~\cite{nextcloud}, Gitea~\cite{gitea}, and Mastodon~\cite{mastodon}.
	The descriptions and statistics are shown in \autoref{tab:extra-benchmarks}.
	For each application, we wrote LLM scripts to generate normal logs. For abnormal logs, we manually injected attacks into normal logs.
	For each application, we wrote 5 scripts to generate abnormal logs and execute each of them several times to generate enough logs.
	Due to limitation of space, detailed attack scenarios are not included in this paper, but can be found at \cite{mines-website}.
	\autoref{tab:overall-eval-more-res} show the evaluation results of \techname on these applications.
	\techname achieves also achieves 100\% precision and high recall on all three applications, indicating that \techname generalizes well to popular web applications.


	\finding{\textbf{RQ5}: \techname generalizes well to popular web applications, achieving high precision and recall on NextCloud, Gitea, and Mastodon.}

%% file: floats/4-number.tex
\begin{table}
	\centering
	\caption{Test \& Attack Scripts Statistics (\# denotes the number of)}
	\mytablefontsize
	\label{tab:script-statistics}
	\begin{tabular}{l|c|c}
		\toprule
		                     & \trainticket & \nicefish \\
		\midrule
		\#Status             & 21          & 13       \\
		\#Normal Operations  & 25          & 27       \\
		\midrule
		\#Test Target APIs   & 48          & 26       \\
		\#Attack Scripts     & 25          & 6        \\
		\#Attack Target APIs & 14          & 13       \\
		\bottomrule
	\end{tabular}
\end{table}

\begin{table}
	\caption{Number of Logs and Attacks}
	\label{tab:dataset-statistics}
	\mytablefontsize
	\begin{tabular}{l|c|c|c|c}
		\toprule
		       & \multicolumn{2}{c|}{\trainticket} & \multicolumn{2}{c}{\nicefish}                      \\
		\midrule
		       & \#Logs                           & \#Attacks                    & \#Logs & \#Attacks \\
		\midrule
		\logsn & 168,799                          & 0                            & 40,654 & 0         \\
		\logln & 8,007                            & 0                            & 3,720  & 0         \\
		\logsa & 6,054                            & 125                          & 4,102  & 120       \\
		\logia & 8,007                            & 125                          & 3,720  & 120       \\
		\bottomrule
	\end{tabular}
\end{table}

%% file: floats/4-overall.tex
\begin{table}
	\centering
	\caption{Overall evaluation of \techname}
	\mytablefontsize
	\label{tab:overall-eval}
	\begin{tabular}{l|c|c|c|c}
		\toprule
		\textbf{Model}                    & \multicolumn{2}{c|}{\textbf{\trainticket}} & \multicolumn{2}{c}{\textbf{\nicefish}}                                        \\
		\midrule
		                                  & \textbf{Precision}                         & \textbf{Recall}                        & \textbf{Precision} & \textbf{Recall} \\
		\midrule
		LogRobust~\cite{zhang2019robust}  & 0.120                                      & 0.650                                  & 0.207              & 0.540           \\
		LogFormer~\cite{guo2024logformer} & 0.272                                      & 0.764                                  & 0.301              & 0.702           \\
		WebNorm~\cite{liao2024detecting}  & \textbf{1.000}                             & 0.704                                  & \textbf{1.000}     & 0.750           \\
		\textbf{\techname (Ours)}         & \textbf{1.000}                             & \textbf{0.948}                         & \textbf{1.000}     & \textbf{0.917}  \\
		\bottomrule
	\end{tabular}
\end{table}

%% file: floats/4-two-types.tex
\begin{table}
	\centering
	\caption{Performance on Two types of Attacks}
	\label{tab:two-type-attacks}
	\mytablefontsize
	\begin{tabular}{l|c|c}
		\toprule
		\textbf{Attack Type}  & \textbf{\trainticket} & \textbf{\nicefish} \\
		\midrule
		\logsa                & 0.896                 & 0.834              \\
		\logia                & 1.000                 & 1.000              \\
		\midrule
		Overall               & 0.948                 & 0.917              \\
		\bottomrule
	\end{tabular}
\end{table}

%% file: floats/4-timecost.tex
\begin{table}
    \centering
    \caption{Performance evaluation of \techname}
    \mytablefontsize
    \label{tab:performance-eval}
    \begin{tabular}{l|c|c}
        \toprule
                                            & \textbf{\trainticket} & \textbf{\nicefish} \\
        \midrule
        \textbf{Number of APIs}             & 48                    & 26                 \\
        \textbf{Training Time (s)}          & 644                   & 406                \\
        \textbf{Training LLM Cost (USD)}    & 20.6                  & 13.7               \\
        \textbf{Running Throughput (log/s)} & $4.0 \times 10^5$     & $2.4 \times 10^5$  \\
        \bottomrule
    \end{tabular}
\end{table}

%% file: floats/4-ablation.tex
\begin{table}
	\centering
	\caption{Ablation Study}
	\mytablefontsize
	\label{tab:ablation-study}
	\begin{tabular}{l|c|c}
		\toprule
		\textbf{Model}                & \textbf{\trainticket} & \textbf{\nicefish} \\
		\midrule
		\textbf{Original (\techname)} & \textbf{0.948}        & \textbf{0.917}     \\
		w/o API-DB relationships      & 0.820                 & 0.834              \\
		w/o API-API relationships     & 0.908                 & 0.836              \\
		w/o API-Env relationships     & 0.780                 & 0.750              \\
		\midrule
		w/o binlog history tracking   & 0.884                 & 0.917              \\
		\midrule
		inducing from raw logs     & 0.896                 & 0.750              \\
		\midrule
		WebNorm                       & 0.704                 & 0.750              \\
		\bottomrule
	\end{tabular}
\end{table}

%% file: floats/4-input-tokens.tex
\begin{table}
	\centering
	\caption{Comparing Input Token Numbers between Raw Log Input and Schema Input}
	\label{tab:comparing-input-token-numbers}
	\mytablefontsize
	\resizebox{\columnwidth}{!}{
		\begin{tabular}{l|l|c|c|c}
			\toprule
			\textbf{Dataset}              & \textbf{Method}              & \textbf{Mean}{}             & \textbf{GeoMean}{}            & \textbf{Median}{}           \\
			\midrule
			\multirow{2}{*}{\trainticket} & Raw Log Input                & $2.40 \times 10^5$          & $4.34 \times 10 ^ 4$          & $2.36 \times 10^4$          \\
			                              & \textbf{Schema Input (Ours)} & $\mathbf{5.01 \times 10^3}$ & $\mathbf{4.86 \times 10 ^ 3}$ & $\mathbf{5.01 \times 10^3}$ \\
			\midrule
			\multirow{2}{*}{\nicefish}    & Raw Log Input                & $1.50 \times 10^4$          & $6.86 \times 10 ^ 3$          & $4.45 \times 10^3$          \\
			                              & \textbf{Schema Input (Ours)} & $\mathbf{3.76 \times 10^3}$ & $\mathbf{4.68 \times 10 ^ 3}$ & $\mathbf{3.54 \times 10^3}$ \\
			\bottomrule
		\end{tabular}
	}
\end{table}

%% file: floats/4-hist-tokens.tex
\begin{figure}
	\centering
	\includegraphics[width=.55\columnwidth]{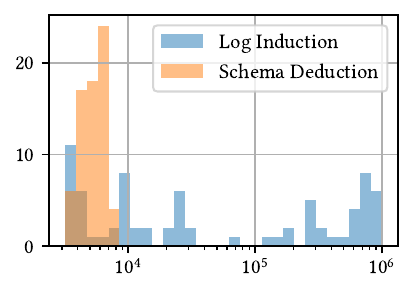}
	\caption{Histogram of the number of tokens in the input logs}
	\Description{
This figure shows a histogram comparing the number of tokens in input prompts between two methods: 
Log Induction and Schema Deduction. The blue bars represent Log Induction, which generally produces 
much longer inputs with token counts spreading up to one million. The orange bars represent Schema 
Deduction, which yields far shorter and more concentrated token lengths around ten thousand or less. 
The figure visually demonstrates that deducing invariants from abstract schemas significantly 
reduces input size compared to inducing from verbose raw logs, improving efficiency for large language models.
	}
	\label{fig:hist-tokens}
\end{figure}

%% file: floats/5-refinement.tex
\begin{table}
	\centering
	\caption{False Negatives Before and After Refinement}
	\label{tab:inv-refinement}
	\resizebox{.9\columnwidth}{!}{
	\mytablefontsize
	\begin{tabular}{l|c|c}
		\toprule
		                 & \trainticket & \nicefish                                      \\
		\midrule
        Dataset Size & 168,799 & 40,654\\
		\midrule
        Number of False Positives (w/o Refinement) & 33,317 & 36,278\\
        False Positive Rate (w/o Refinement) & 0.197 & 0.892\\
        \midrule
        Number of False Negatives (w/ Refinement) & 0 & 0\\
        False Negative Rate (w/ Refinement) & 0.000 & 0.000\\
		\bottomrule
	\end{tabular}
	}
\end{table}

%% file: floats/4-different-llms.tex
\begin{table}
	\centering
	\caption{Comparison of Different LLMs}
	\label{tab:comparing-llms}
	\mytablefontsize
	\begin{tabular}{l|c|c|c|c}
		\toprule
		            & \multicolumn{2}{c|}{\textbf{\trainticket}} & \multicolumn{2}{c}{\textbf{\nicefish}}                                        \\
		\cmidrule{2-5}
		            & \textbf{Percision}                         & \textbf{Recall}                        & \textbf{Percision} & \textbf{Recall} \\
		\midrule
		GPT-4o      & 1.000                                      & \textbf{0.948}                         & 1.000              & \textbf{0.833}  \\
		GPT-4o-mini & 1.000                                      & 0.800                                  & 1.000              & 0.750           \\
		Claude 3.7  & 1.000                                      & 0.888                                  & 1.000              & \textbf{0.833}  \\
		DeepSeek-V3 & 1.000                                      & 0.936                                  & 1.000              & \textbf{0.833}  \\
		\bottomrule
	\end{tabular}
\end{table}

%% file: floats/5-naming.tex
\begin{table}
	\centering
	\caption{Impact of Naming Conventions on \techname}
	\mytablefontsize
	\label{tab:naming-conventions}
	\begin{tabular}{l|c|c}
		\toprule
		                                             & Precision & Recall \\
		\midrule
		Original Names (e.g., \texttt{QueryByBatch}) & 1.00      & 0.94   \\
		Snake Case (e.g., \texttt{query\_by\_batch}) & 1.00      & 0.94   \\
		Camel Case (e.g., \texttt{QueryByBatch})     & 1.00      & 0.94   \\
		Concat All (e.g., \texttt{querybybatch})     & 1.00      & 0.94   \\
		Partial Abbreviation (e.g., \texttt{query})  & 1.00      & 0.93   \\
		Extreme Abbreviation (e.g., \texttt{q})      & 1.00      & 0.84   \\
		\bottomrule
	\end{tabular}
\end{table}

%% file: floats/4-more-benchmarks.tex
\begin{table*}
    \centering
    \caption{Extra benchmarks to evaluate generalization of \techname}
    \mytablefontsize
    \label{tab:extra-benchmarks}
    \resizebox{\textwidth}{!}{
    \begin{tabular}{l|c|c|c|c|c|c|c}
        \toprule
        \textbf{Project}      & Line of Code & Language & Web Framework & Database   & GitHub Stars & Number of Microservices & Number of Database Tables \\
        \midrule
        \textbf{\trainticket} & 37.8k        & Java     & Spring        & MySQL      & 804          & 41                      & 34                        \\
        \textbf{\nicefish}    & 4.7k         & Java     & Spring        & MySQL      & 732          & 2                       & 16                        \\
        \midrule
        \textbf{\gitea}       & 330.7k       & Go       & Gin           & MySQL      & 49.6k        & 1                       & 114                       \\
        \textbf{\mastodon}    & 109.8k       & Ruby     & Ruby on Rails & PostgreSQL & 48.6k        & 1                       & 98                        \\
        \textbf{\nextcloud}   & 426.0k       & PHP      & Vanilla PHP   & MariaDB    & 30.2k        & 1                       & 129                       \\
        \bottomrule
    \end{tabular}
    }
\end{table*}

%% file: floats/4-more-benchmarks-res.tex
\begin{table}
    \centering
    \caption{Overall evaluation of \techname on extra benchmarks}
    \mytablefontsize
    \label{tab:overall-eval-more-res}
    \resizebox{\columnwidth}{!}{
        \begin{tabular}{l|l|c|c|c|c}
            \toprule
            \textbf{Benchmark} & \textbf{Metric} & \textbf{LogRobust~\cite{zhang2019robust}} & \textbf{LogFormer~\cite{guo2024logformer}} & \textbf{WebNorm~\cite{liao2024detecting}} & \textbf{\techname (Ours)} \\
            \midrule
            \multirow{3}{*}{Gitea}
                               & Precision       & 0.640                                     & 0.481                                      & \textbf{1.000}                            & \textbf{1.000}            \\
                               & Recall          & \textbf{0.970}                            & 0.474                                      & 0.176                                     & \textbf{0.956}            \\
                               & F1              & 0.771                                     & 0.477                                      & 0.299                                     & \textbf{0.977}            \\
            \midrule
            \multirow{3}{*}{Mastodon}
                               & Precision       & 0.233                                     & 0.454                                      & \textbf{1.000}                            & \textbf{1.000}            \\
                               & Recall          & \textbf{1.000}                            & 0.625                                      & 0.667                                     & \textbf{0.833}            \\
                               & F1              & 0.377                                     & 0.526                                      & 0.800                                     & \textbf{0.909}            \\
            \midrule
            \multirow{3}{*}{NextCloud}
                               & Precision       & 0.064                                     & 0.059                                      & \textbf{1.000}                            & \textbf{1.000}            \\
                               & Recall          & \textbf{1.000}                            & 0.300                                      & 0.750                                     & \textbf{0.906}            \\
                               & F1              & 0.120                                     & 0.099                                      & 0.857                                     & \textbf{0.950}            \\
            \bottomrule
        \end{tabular}
    }
\end{table}

%% file: chaps/05-discussion.tex
\section{Discussion and Limitations}

Despite the strong performance of \techname across benchmarks, several limitations remain.

\noindent\textbf{Migration Complexity.}
Adapting \techname to new systems may require engineering effort, including handling diverse database backends, session management, and API endpoints. Our framework minimizes technology dependencies by relying only on common components such as APIs and database schemas.

\noindent\textbf{Dependency on Naming Quality.}
The effectiveness of \techname depends on meaningful and consistent naming in APIs and schemas. While this assumption generally holds, systems with obfuscated or inconsistent naming remain challenging.

\noindent\textbf{Optional but Beneficial Documentation.}
The current implementation does not utilize application-level documentation. Although not required, documentation such as ER diagrams or type annotations can significantly improve invariant inference. For example, on \trainticket, \techname failed to detect an attack where the API field \texttt{seatClass} was set to 5, while valid values were only 0 and 1. Adding a single line of documentation, ``\textit{valid values for seat class are 0 and 1}'', enabled \techname to synthesize the correct invariant and detect the attack. This suggests that even minimal documentation can enhance semantic precision and motivates documentation-aware extensions.

\noindent\textbf{Requirement for Comprehensive Logs.}
Comprehensive logs are essential for capturing representative behaviors and reducing false positives. While typically available in testing or staging environments, performance may degrade in log-sparse scenarios.

%% file: chaps/06-related-work.tex
\section{Related Work}

\noindent\textbf{Log Anomaly Detection.}
Log anomaly detection can be traced back to execution trace analysis. Some works focus on analyzing the execution of specific APIs or methods to find anomalies \cite{ye2024spurious}.
Traditional log anomaly detection approaches rely on predefined rules \cite{hansen1993automated,oprea2015detection,prewett2003analyzing,rouillard2004real,roy2015perfaugur,yamanishi2005dynamic,yen2013beehive}, which are limited to specific application scenarios and require domain expertise \cite{du2017deeplog}.

In recent years, researchers have proposed learning-based approaches to automatically learn normal behaviors from logs \cite{acharya2007mining, lorenzoli2008automatic, walkinshaw2008inferring, pradel2009automatic, beschastnikh2011leveraging, krka2014automatic, breier2015anomaly, amar2018using,rufino2020improving,stocco2020towards,kang2019spatiotemporal,njoku2025kernel,wu2023effectiveness,lupton2021literature,alam2019framework,schneider2010synoptic},
which can be categorized into two types.

The first category utilizes neural networks to directly predict whether a log sequence is normal or anomalous \cite{du2017deeplog, goldstein2017experience, brown2018recurrent, lu2018detecting, meng2019loganomaly, huang2020hitanomaly, li2020swisslog, yuan2020ada, wang2021multi, le2021log, yang2021plelog, guo2021logbert, fu2023mlog, zhao2021identifying, han2023loggpt, qi2023loggpt, tao2023biglog, el2024replicawatcher, zhang2019robust, guo2024logformer, almodovar2024logfit}. These methods utilize different neural network architectures to enhance the performance of web anomaly detection, including RNNs \cite{du2017deeplog, brown2018recurrent}, CNNs \cite{lu2018detecting, fu2023mlog}, Transformers \cite{huang2020hitanomaly, guo2024logformer}, GNNs \cite{zhang2022deeptralog}, pretrained language models \cite{guo2021logbert, han2023loggpt}, and instrumented large language models \cite{qi2023loggpt}. Some works also utilize unsupervised or semi-supervised learning to alleviate the need for labeled data \cite{yang2021plelog, meng2019loganomaly}. These methods can capture the temporal dependencies in log sequences and improve the performance of log anomaly detection. However, they often lack explainability in their detection results and may struggle to capture subtle changes in abnormal logs.

The second category focuses on learning explainable normalities to detect anomalies \cite{liao2024detecting}. The only work in this category is WebNorm \cite{liao2024detecting}. WebNorm detects web anomalies by learning normality first-order logic rules for web applications. This method offers better explainability and can capture subtle but crucial changes. However, WebNorm only focuses on analyzing web logs and does not consider the relational integrity between web logs and the underlying database. Our proposed method, \techname, aims to address these limitations by inspecting the relational integrity between web logs and the database and generating normality rules based on the abstract schema and the ER diagram.

\noindent\textbf{RESTful API Security.}
RESTful APIs employ a stateless architecture and standard HTTP methods, and are now widely used in web applications, making their security a major concern. 
Numerous studies have addressed RESTful API security, primarily through automated test case generation to detect vulnerabilities~\cite{deng2023nautilus, du2024vulnerability, atlidakis2019restler, viglianisi2020resttestgen, martin2020restest,martin2020automated,xu2021chunk}. These approaches typically fuzz API sequences and insert attack or detection invocations to uncover issues, leveraging API specifications, data dependencies~\cite{viglianisi2020resttestgen, martin2020restest}, and neural network predictions~\cite{lyu2023miner,xu2023improving}.
Some works further enhance testing via targeted fuzzing strategies~\cite{deng2023nautilus, du2024vulnerability}. These methods are mainly designed to reveal vulnerabilities such as SQL injection and cross-site scripting. 

In contrast, our work aims to generate normality rules for RESTful APIs, strengthening web application security by specifically detecting attacks that violate the intended normal behaviors of web applications.

%% file: chaps/07-conclusion.tex
\section{Conclusion}

In this paper, we propose \techname, a novel rule-learning based approach to enhance web application security.
By leveraging \textit{deducing from specifications} and utilizing information beyond log instrumentation, \techname is more effective in detecting anomalies in web applications.
Experiments on two datasets demonstrate that \techname outperforms existing approaches.








